\documentclass[mathleft
]{an}

\usepackage{graphicx}
\usepackage{times}
\begin{document}

\Pagespan{789}{}
\Yearpublication{2006}%
\Yearsubmission{2005}%
\Month{11}%
\Volume{999}%
\Issue{88}%

\title{
Magnetic field studies of massive main sequence stars
}

\author{M.~Sch\"oller\inst{1}\fnmsep\thanks{Corresponding author:
  \email{mschoell@eso.org}\newline}
\and S.~Hubrig\inst{2}
\and I.~Ilyin\inst{2}
\and N.~V.~Kharchenko\inst{2,3}
\and M.~Briquet\inst{4}
\and N.~Langer\inst{5}
\and L.~M.~Oskinova\inst{6}
\and the MAGORI collaboration
}
\titlerunning{
Magnetic field studies of massive main sequence stars
}
\authorrunning{
M.~Sch\"oller et al.
}
\institute{
European Southern Observatory, Karl-Schwarzschild-Str.~2, 85748~Garching, Germany
\and
Leibniz-Institut f\"ur Astrophysik Potsdam (AIP), An der Sternwarte 16, 14482 Potsdam, Germany
\and 
Main Astronomical Observatory, 27 Academica Zabolotnogo Str., 03680 Kiev, Ukraine
\and 
Instituut voor Sterrenkunde, Katholieke Universiteit Leuven, Celestijnenlaan 200 D, 3001, Leuven, Belgium
\and 
Argelander-Institut f\"ur Astronomie, Universit\"at Bonn, Auf dem H\"ugel~71, 53121~Bonn, Germany
\and 
Universit\"at Potsdam, Institut f\"ur Physik und Astronomie, 14476~Potsdam, Germany
}

\received{30 December 2011}
\accepted{31 December 2011}
\publonline{later}


\abstract{%
We report on the status of our spectropolarimetric observations of massive stars.
During the last years, we have discovered magnetic fields in many objects
of the upper main sequence, including Be stars, $\beta$\,Cephei and Slowly
Pulsating B stars, and a dozen O stars.
Since the effects of those magnetic
fields have been found to be substantial by recent models, we are looking into
their impact on stellar rotation, pulsation, stellar winds, and chemical abundances.
Accurate studies
of the age, environment, and kinematic characteristics of the magnetic stars are
also promising to give us new insight into the origin of the magnetic fields.
Furthermore, longer time series of magnetic field measurements allow us to
observe the temporal variability of the magnetic field and to deduce the stellar
rotation period and the magnetic field geometry.
Studies of the magnetic field
in massive stars are indispensable to understand the conditions controlling the
presence of those fields and their implications on the stellar physical
parameters and evolution.
  }

\maketitle

\section{Introduction}
\label{sect:intro}

Magnetic fields play an important role in astrophysical phenomena of the universe on various scales.
For galaxies, dynamo models associated with various MHD instabilities occurring in the interstellar medium (ISM)
are used  to explain the formation of the galactic structure (e.g., Gomez \& Cox \cite{GomezCox2004};
Bonanno \& Urpin \cite{BonannoUrpin2008}).
Magnetic fields play a role in the evolution of interstellar molecular clouds and
the star formation process, where
the cloud collapse is probably taking place along the magnetic field lines (e.g.\ Alves et al.\ \cite{Alves2008}).
They are also present at all stages of stellar evolution,
from young T\,Tauri stars and Ap/Bp stars
to the end products: white dwarfs, neutron stars, and magnetars.
On the other hand, the role of magnetic fields
in massive O-type stars and Wolf-Rayet (WR) stars remains unknown.
No definitive magnetic field has ever been detected in
WR stars and presently only about a dozen O stars have published magnetic fields, while
for early B stars magnetic field detections are more numerous.
In our study of the open cluster NGC\,3766, we found that seven out of the 14 observed early B-type stars 
showed magnetic fields (Hubrig et al.\ \cite{Hubrig2009}).
In addition, theories about the origin of magnetic fields in O- and early B-type stars
remain poorly developed, mostly because the distribution of magnetic field strengths in massive stars
from the ZAMS to more evolved stages has not yet been studied.
In our recent studies, we focused on magnetic fields of massive stars observed in different environments:
in open clusters at different ages and in the field (Hubrig et al.\ \cite{Hubrig2009,Hubrig2011c,Hubrig2011e}).


\section{Magnetic field models for $\beta$\,Cephei and SPB stars}
\label{sect:earlyB}


\begin{figure*}
\centering
\includegraphics[angle=270,totalheight=0.32\textwidth]{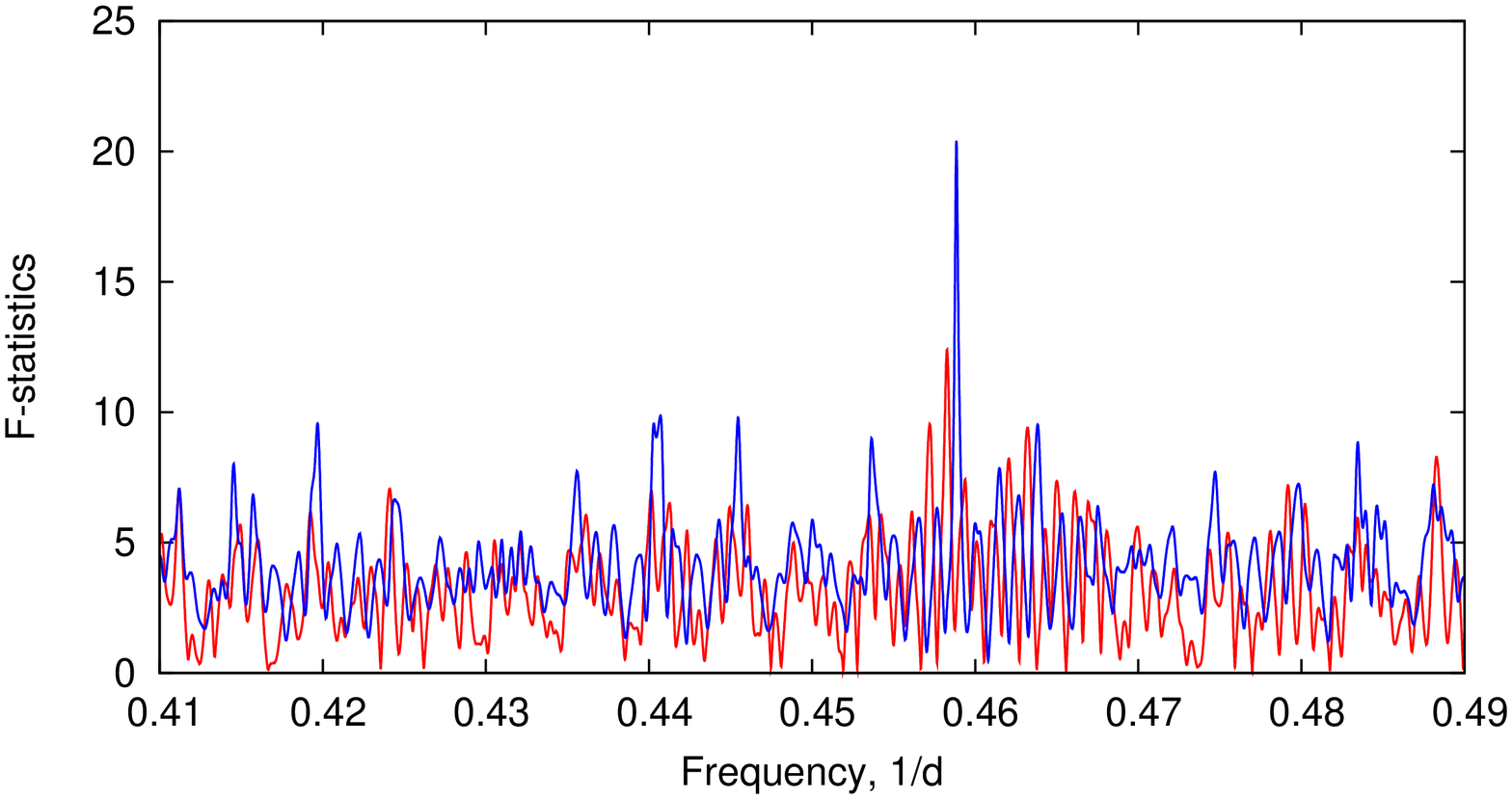}
\includegraphics[angle=270,totalheight=0.32\textwidth]{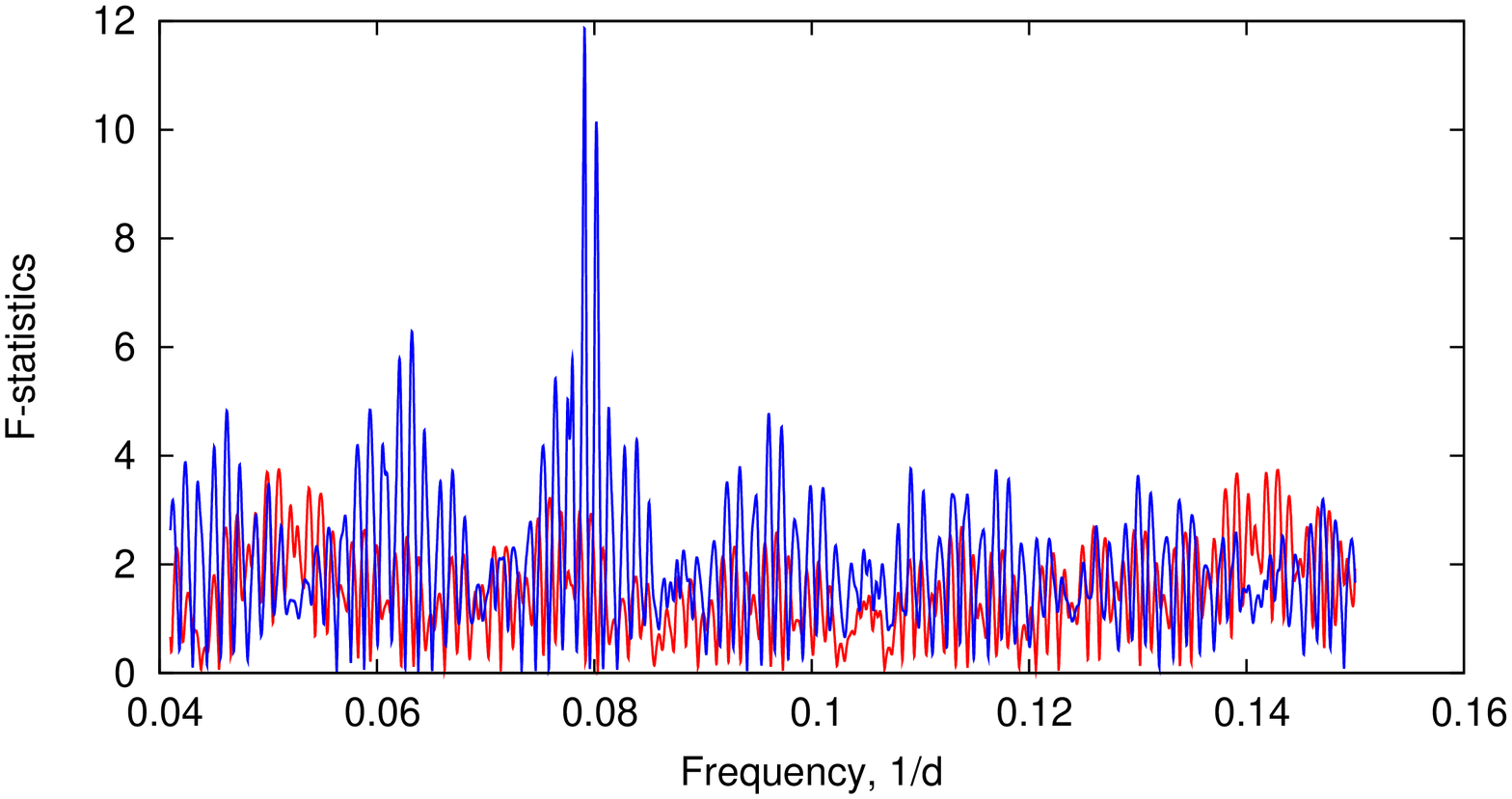}
\includegraphics[width=0.45\textwidth]{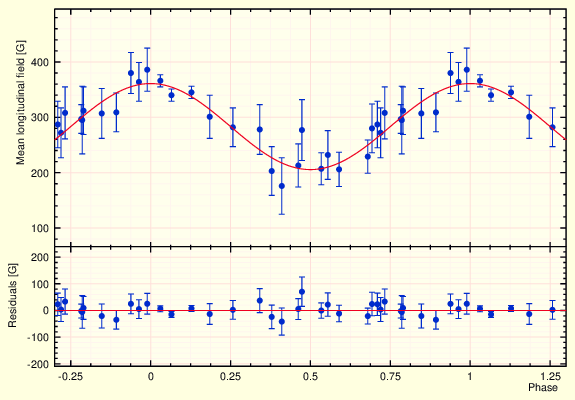}
\includegraphics[width=0.45\textwidth]{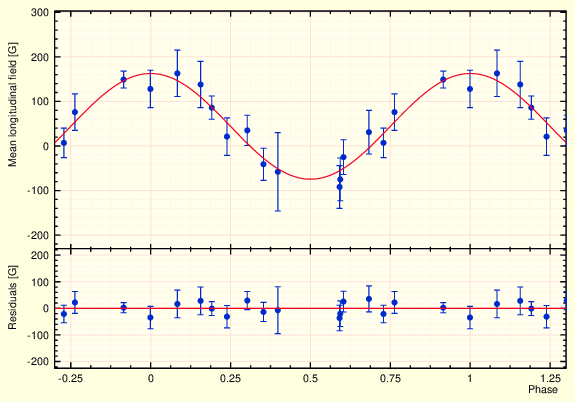}
\caption{
Periodograms (top) and phase diagrams (bottom) with the best sinusoidal fit for
the longitudinal magnetic field measurements of $\xi^1$\,CMa (left) and 15\,CMa (right).
The residuals (Observed $-$ Calculated) are shown in the lower panels.
}
\label{fig:betcep_fits}
\end{figure*}

Using FORS\,1/2 and SOFIN longitudinal magnetic field measurements collected in
our recent studies (Hubrig et al.\ \cite{Hubrig2011a}), we were able to determine the rotation period and constrain
the field geometry of two $\beta$\,Cephei stars, one candidate $\beta$\,Cephei star, and one SPB star.
The dipole model provides a satisfactory fit to the data and among the very few
presently known magnetic $\beta$\,Cephei stars, $\xi^1$\,CMa and $\alpha$\,Pyx
possess with several kG the largest dipole strengths.
In Fig.~\ref{fig:betcep_fits} we show periodograms and phase diagrams
with the best sinusoidal fit for the longitudinal magnetic field measurements of
$\xi^1$\,CMa (left) and 15\,CMa (right).

\begin{figure}
\centering
\includegraphics[width=0.45\textwidth]{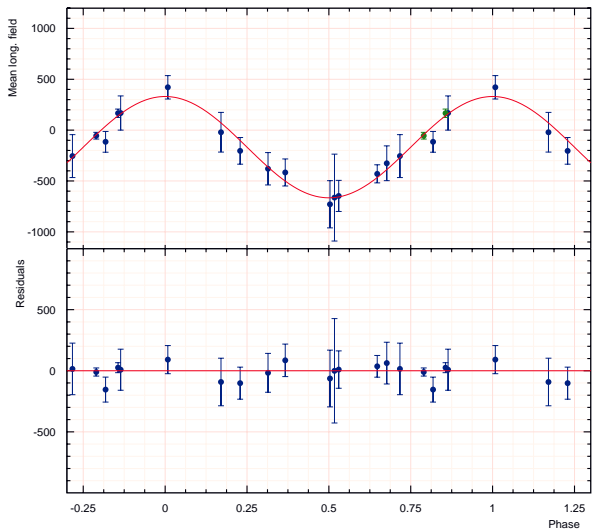}
\caption{
Periodogram for the magnetic field measurements of V1449\,Aql.
The residuals (Observed $-$ Calculated) are shown in the lower panel.
}
\label{fig:v1449aqla}
\end{figure}

\begin{figure}
\centering
\includegraphics[width=0.45\textwidth]{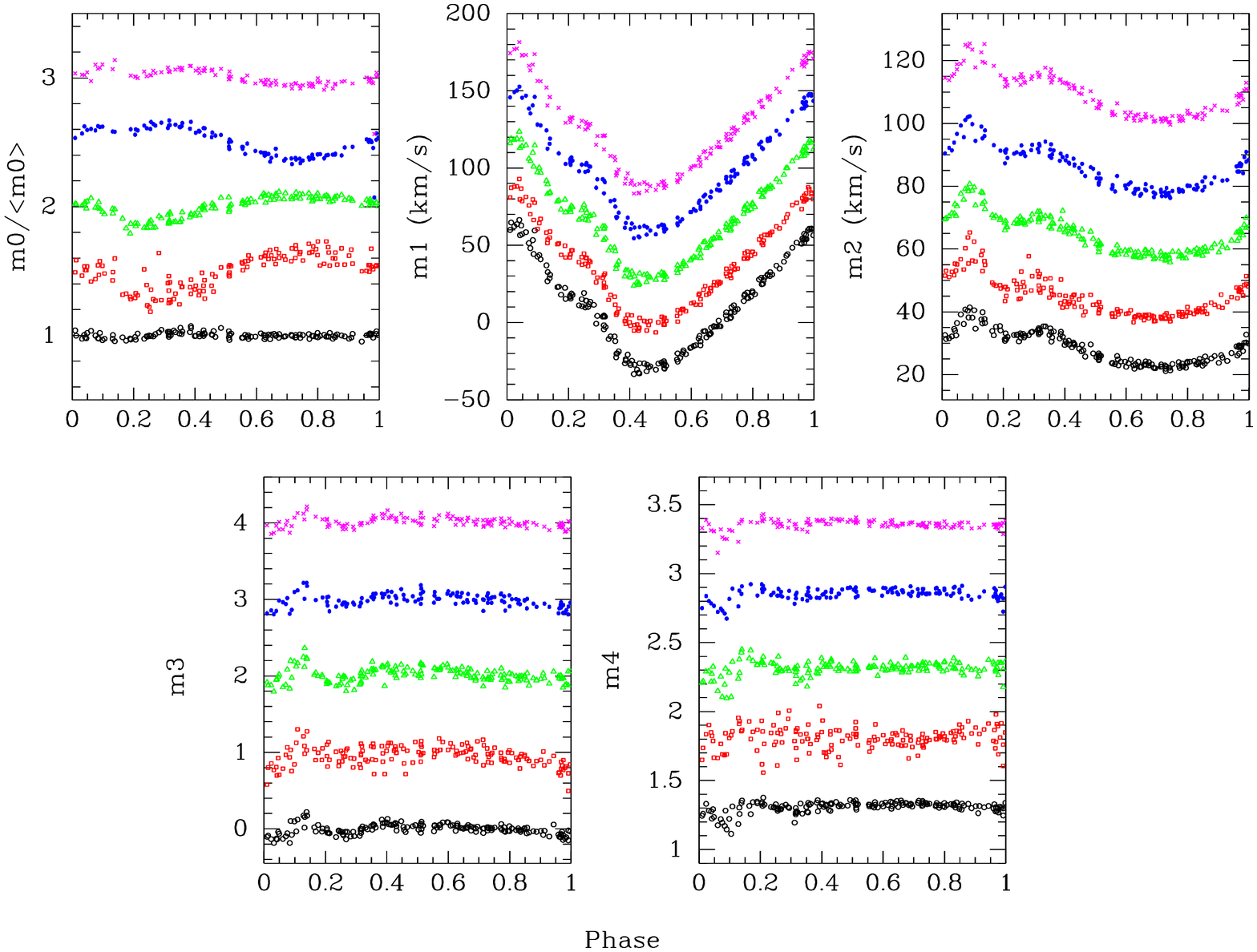}
\caption{
Equivalent width (m0), radial velocity (m1), line width (m2), asymmetry (m3), and kurtosis (m4)
from our SOFIN data of V1449\,Aql.
}
\label{fig:v1449aqlb}
\end{figure}

In our studies,
V1449\,Aql possesses so far the strongest longitudinal magnetic field of up to 700\,G among the $\beta$\,Cephei stars.
The resulting periodogram from our SOFIN and FORS1\, measurements displays three
dominant peaks with the highest peak at $f=0.0720$\,d$^{-1}$ corresponding to a period $P=13.893$\,d 
(Hubrig et al.\ \cite{Hubrig2011b}).
This period was recently confirmed with seismic modeling based on CoRoT data
and ground-based time-resolved spectroscopy by Aerts et al.\ (\cite{Aerts2011}),
who also mention a somewhat high initial internal metallicity in their models,
probably related to the presence of a magnetic field.
The magnetic field geometry can likely be described by a centered dipole with a polar magnetic field strength
$B_{\rm d}$ around 3\,kG.
In Fig.~\ref{fig:v1449aqla} we show the periodogram for V1449\,Aql and the residuals for our measurements.
The line profiles undergo very significant pulsational variability and any
rotational modulation of the observed profiles is completely masked by the pulsational modulation.
The moment variations for five lines belonging to different elements, namely
He~{\sc i} 4713.1, C~{\sc ii} 5133.1, N~{\sc ii} 5667.6, O~{\sc ii} 4592.0, and Si~{\sc iii} 4553.6
exhibit slight differences between the elements.
In Fig.~\ref{fig:v1449aqlb} we show the equivalent width (m0),
radial velocity (m1), line width (m2), asymmetry (m3), and kurtosis (m4) from our SOFIN data of V1449\,Aql.


\section{O-type stars}
\label{sect:o}


\begin{figure}
\centering
\includegraphics[width=0.45\textwidth]{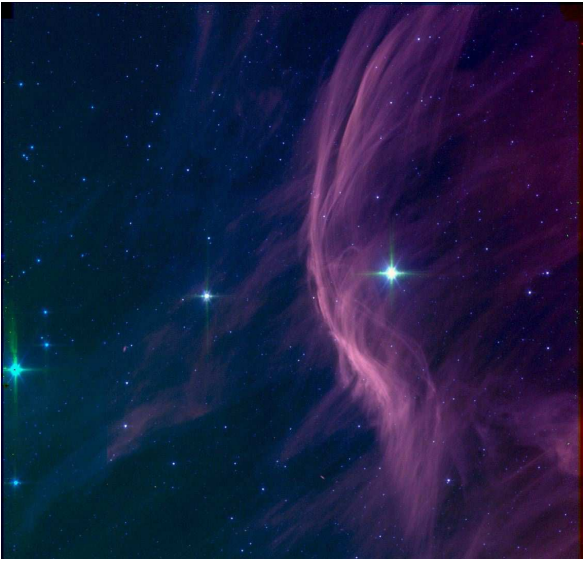}
  \caption{
Combined IR Spitzer IRAC image of the bow shock around $\zeta$\,Oph.
}
\label{fig:zoph_spitzer}
\end{figure}


To investigate statistically whether magnetic fields in O-type stars are
ubiquitous or appear only in stars with a specific spectral classification,
certain age, or in a special environment, we acquired new spectropolarimetric observations with FORS\,1/2.
A field at a significance level of 3$\sigma$ was detected in eleven stars (Hubrig et al.\ \cite{Hubrig2011c,Hubrig2011d}).
The strongest longitudinal magnetic fields were measured in two Of?p stars:
$\left<B_{\rm z}\right>=−381\pm122$\,G for CPD--28\,2561 and $\left<B_{\rm z}\right>=−297\pm62$\,G for HD\,148937.
Both magnetic fields were detected by us for the first time, the latter in an earlier study
(Hubrig et al.\ \cite{Hubrig2008}).

The star $\zeta$\,Ophiuchi of spectral type O9.5\,V is a
well-known rapidly rotating runaway  star with extremely interesting
characteristics. It undergoes episodic mass loss seen as emission in
H$\alpha$, and it  is possible that it rotates with almost break-up
velocity with $v$\,sin\,$i=400$\,km\,s$^{-1}$ (Kambe et al.\ \cite{Kambe1993}).
Our spectropolarimetric observations 
of $\zeta$\,Oph with FORS\,1 in 2008 revealed the presence of a mean longitudinal magnetic field 
$\left< B_z\right>_{\rm all}= 141\pm45$\,G. Additional 
nine spectropolarimetric observations were obtained with FORS\,2 over the rotation period in 2011. 
The longitudinal 
magnetic field shows a change of polarity and its variation over the rotation cycle 
can be represented  by a sinusoidal fit with a semi-amplitude of $\sim$160\,G.
Fig.~\ref{fig:zoph_spitzer} shows a combined IR {\em Spitzer} IRAC image of the bow shock around $\zeta$\,Oph.

\begin{figure}
\centering
\includegraphics[width=0.45\textwidth]{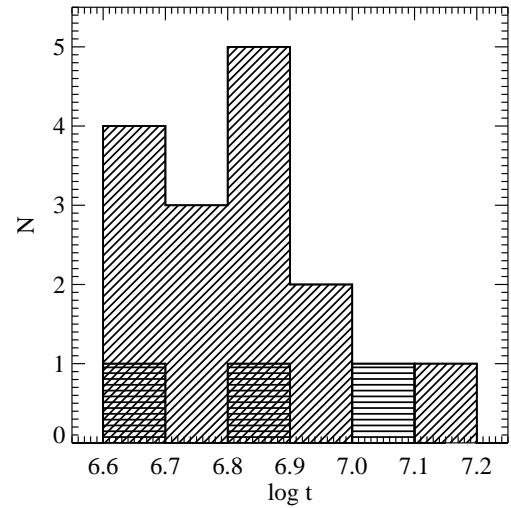}
  \caption{
Age distribution of probable cluster members from our study.
Stars with magnetic fields are denoted by horizontal lines.
}
\label{fig:histo}
\end{figure}

The available observations (Hubrig et al.\ \cite{Hubrig2011c,Hubrig2011e}) seem to indicate that a magnetic field is more
frequently detected in field stars than in stars belonging to clusters or associations.
It is striking that most previously detected magnetic O-type stars are candidate runaway stars.
Clearly, these findings generate a strong motivation to carry out a kinematic
study of all stars previously surveyed for magnetic fields to search for a
correlation between the kinematic status and the presence of a magnetic field.
In Fig.~\ref{fig:histo} we show the age distribution of probable cluster members from our study.
The three stars with magnetic fields, HD\,155806, HD\,156154, and HD\,164794, are denoted by horizontal lines.

\begin{figure}
\centering
\includegraphics[width=0.45\textwidth]{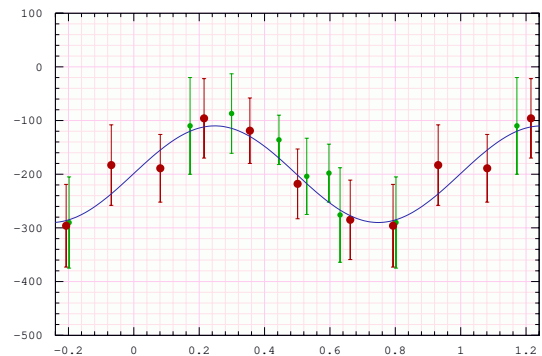}
\caption{
Longitudinal magnetic field variation of HD\,148937 over the 7.032\,d period
determined by Naz\'e et al.\ (\cite{Naze2010}).
Red (large) symbols correspond to ESPaDOnS observations (Wade et al.\ \cite{Wade2011}),
while green (small) symbols are our FORS\,1 and FORS\,2 measurements
(Hubrig et al.\ \cite{Hubrig2008,Hubrig2011c}; Hubrig et al.\ in preparation).
}
\label{fig:hd148}
\end{figure}

To demonstrate the excellent potential of FORS\,2 for the
detection and investigation of magnetic fields in massive stars, we present in Fig.~\ref{fig:hd148} 
our FORS\,2 observations collected between 2008 and May 2011 of the
Of?p star HD\,148937, together with the ESPaDOnS observations obtained at the CFHT
(Wade et al.\ \cite{Wade2011}).
The measurement errors for both ESPaDOnS and FORS\,1/2 observations
are of similar order.
One more measurement (not shown in this Figure) was
obtained in 2010 on May 22, but ordinary and extraordinary beams in the FORS\,2
setup were overlapping in this exposure due to slitlet problems.

\section{Discussion}
\label{sect:disc}

Early B-type pulsating stars had been a puzzle for a long time with
respect to the presence of magnetic fields in their atmospheres.
Only very few stars of this type with very weak magnetic fields were detected
before we started our surveys of magnetic fields in B-type stars in 2004.
We were the first to determine magnetic field strengths for a large sample of O-type stars, with an accuracy of a few tens of Gauss.
Very few magnetic fields stronger than 300\,G were detected in the studied sample,
suggesting that large-scale, dipole-like magnetic fields with polar magnetic
field strengths higher than 1\,kG are not widespread among O-type stars.
Our studies of massive stars revealed that the presence of a magnetic field can
be expected in stars of different classification categories and at different evolutionary stages.
We note that no physical properties are known that define these particular classes of stars as non-magnetic.
The inability to detect magnetic fields in massive stars in previous studies
could be related to the weakness of these fields, which can, in some stars, be as little as only a few tens of Gauss.

The results of our kinematic analysis of known magnetic O-type stars using the best available
astrometric, spectroscopic, and photometric data indicates that a magnetic field
is more frequently detected in candidate runaway stars than in stars belonging to
clusters or associations (Hubrig et al.\ \cite{Hubrig2011e}).
As the sample of stars with magnetic field detections is still very small,
a study of a larger sample is urgently needed to confirm the detected trend by performing dedicated
magnetic field surveys of O stars in clusters/associations and in the field.
The results obtained so far allow us to constrain preliminarily the conditions
conducive to the presence of magnetic fields and derive the first trends for their occurrence rate
and field strength distribution. This information is critical for answering the principal question of
the possible origin of magnetic fields in massive stars.

\end{document}